\input{aipcheck}

\documentclass[
 ,final            % use final for the camera ready runs
  ]
  {aipproc}

\layoutstyle{6x9}

\def\di{\displaystyle}

\def\bg{\begin{eqnarray}\begin{array}{rcl}\displaystyle}
\def\eg{\end{array} &\di    &\di   \end{eqnarray}}
\def\bm#1{\begin{eqnarray}\begin{array}{#1}\di}
\def\bmo#1{\begin{eqnarray*}\begin{array}{#1}\di}
\def\bml#1#2{\begin{eqnarray}\begin{array}{#1}\label{#2}\di}
\def\bgo{\begin{eqnarray*}\begin{array}{rcl}\displaystyle}
\def\ego{\end{array} &\di    &\di \nonumber  \end{eqnarray*}}

\def\btensor#1#2{\renew\left#1\begin{array}{#2}\di}
\def\brtensor#1#2#3{\ren#3\left#1\begin{array}{#2}}
\def\botensor#1#2{\renew\left#1\begin{array}{#2}}
\def\etensor#1{\end{array}\right#1}

\def\eq#1{(\ref{#1})}
\def\Eq#1{Eq.~(\ref{#1})}

\def\s0#1#2{\mbox{\small{$ \frac{#1}{#2} $}}}
\def\0#1#2{\frac{#1}{#2}}

\def\ren#1{\renewcommand{\arraystretch}{#1}}

\def\renew{\renewcommand{\arraystretch}{1}}
\begin{document}
%preprint{FAU-TP3-04-05}
%preprint{HD-THEP-04-55}
\title{Signatures of confinement in Landau gauge QCD}

\classification{05.10.Cc, 11.15.Tk, 12.38.Aw}
\keywords      {confinement, gluon propagator, ghost propagator, flow equation}

\author{J.~M. Pawlowski
}{address={Institute for Theoretical 
Physics, University of 
Heidelberg, 69120 Heidelberg, Germany}
}

\author{D.~F.~Litim}{
  address={Theory Group, CERN, CH-1211 Geneva 23; 
SPA, U. Southampton, Southampton SO17 1BJ, U.K.}
}

\author{S.~Nedelko}{
  address={BLTP, JINR, 141980 Dubna, Russia}
}

\author{L.~von Smekal}{address={Institute for Theoretical 
Physics III, 
Universit\"at Erlangen, D-91058 Erlangen, Germany}
}

\begin{abstract}
We summarise an analysis of the infrared regime of Landau gauge QCD by
means of a flow equation approach \cite{Pawlowski:2003hq}. 
The infrared behaviour of gluon and
ghost propagators is evaluated. The results provide further evidence
for the Kugo-Ojima confinement scenario. We also discuss their
relation to results obtained with other functional methods as well as the 
lattice.

\end{abstract}

\maketitle

%%%%%%%%%%%%%%%%%%%%%%%%%%%%%%%%%%%%%%%%%%%%
%% MAINMATTER
%%%%%%%%%%%%%%%%%%%%%%%%%%%%%%%%%%%%%%%%%%%%
In gauge fixed formulations of QCD the infrared behaviour of ghost and
gluon propagators provide signatures of confinement: in Landau gauge
QCD the confinement scenarios of Kugo-Ojima \cite{Kug79} and
Gribov-Zwanziger, e.g.\  \cite{Zwanziger:2001kw}, entail an infrared
enhancement for the ghost propagation and an infrared suppression for
the gluon propagation. This behaviour was first seen within a
functional approach using Dyson-Schwinger equations (DSE)
\cite{Sme97}, giving access to the full momentum regime. It was later
confirmed within lattice studies down to scales about 1 GeV.  However,
in the deep infrared below 1 GeV, lattice studies encounter problems
due to finite size effects. This situation calls for an independent
confirmation of the infrared behaviour seen in DS-studies.  Ideally
such a method would still share enough structure with other functional
methods such as DSEs in order to benefit from insights and results
obtained from these equations. The above features are precisely given
for the flow equation, a particular advantage of which is its
flexibility when it comes to approximations. So far this approach has
been used in Landau gauge QCD for high and intermediate momenta
\cite{Ellwanger:1996qf}.

Here we present results of a flow equation approach to the infrared
regime of Landau gauge QCD \cite{Pawlowski:2003hq}. 
Our analysis is based on an integrated flow equation
that reads for the scale-dependent effective action $\Gamma_k$, 
\begin{equation}\label{eq:flow}
\Gamma_0-\Gamma_k = 
\frac{1}{2}\, \int_0^k d k\,  
{\rm Tr}\,\frac{1}{\Gamma^{(2)}_k+R}\,\partial_k R\,.  
\end{equation}
The scale $k$ is an infrared scale below which $\Gamma_k$ has no
propagating degrees of freedom and $\Gamma^{(n)}$ stands for its $n$th
derivative w.r.t.\ the fields. Consequently $\Gamma=\Gamma_0$ is the
full effective action. Flows for Green functions are obtained by
taking field derivatives of \eq{eq:flow}. The resulting equations
share many features with DSEs and stochastic quantisation for Green
functions. However, in contradistinction to those approaches built on
dressed {\it and} bare quantities, the flow equation \eq{eq:flow}
links dressed vertices and propagators exclusively. Moreover, the flow
equation and its $k$-integral \eq{eq:flow} are manifestly ultraviolet
and infrared finite, no additional renormalisation is required even
within truncations.  Therefore the flow equation offers an interesting
functional method for accessing the infrared regime: its close
relation to DSEs makes many of the truncation schemes and results of
DSEs accessible to flow equation studies; its qualitative differences
and complementary advantages provide additional support for results
obtained within both approaches.

We evaluate the integrated flow \eq{eq:flow} of ghost and gluon 
propagators in the deep infrared 
\begin{eqnarray}\label{eq:IR1}
p^2 \ll \Lambda_{\rm QCD}^2\,,
\end{eqnarray}
where $\Lambda_{\rm QCD}$ is the dynamical mass scale of QCD. For
these momenta the integrated flow \eq{eq:flow} tends to zero as the
flow reaches a (trivial) fixed point at $ k=0$. This enables us to
study the leading infrared behaviour of the propagators by means of a
fixed point argument developed in \cite{Pawlowski:2003hq}.  For
momenta in the regime \eq{eq:IR1} and for $k^2\ll \Lambda_{\rm QCD}^2$
we can expand $n$-point functions at finite $k$ about that at $k=0$
with
\begin{eqnarray}\label{eq:IR2}
\Gamma_k^{(n)}=\Gamma_0^{(n)}(1+\delta Z_n)\,,
\end{eqnarray} 
where $\delta Z_n$ only depend on ratios $p_i/k$ with
$i=1,...,n-1$. \Eq{eq:IR2} is valid up to order $p_i/\Lambda_{\rm
QCD}$. Indeed, \eq{eq:IR2} can be proven by iterating the integrated 
flow \eq{eq:flow} about $\delta Z_n\equiv 0$. So far we have not relied on any 
approximation. For the explicit computation we resort to a truncation with 
dressed vertices with trivial momentum dependence, an 
approximation which is well in accord with consistency considerations 
\cite{Schleifenbaum:2004id} as well as lattice studies 
\cite{Cucchieri:2004sq}. We allow for a general momentum dependence on 
$x=p^2/k^2$ in the ghost and (transversal) gluon two point functions 
$\Gamma_{k,C}^{(2)}$ and $\Gamma_{k,A}^{(2)}$ 
\begin{eqnarray}\nonumber 
\Gamma_{k,C}^{(2)}(x)& \simeq & z_{C}\, p^2\, x^{\kappa_{C}} 
(1+\delta Z_{C}(x))
\,,\\ 
\Gamma_{k,A}^{(2)}(x)& \simeq 
& z_{A}\, p^2\, x^{-2 \kappa_{C}} (1+\delta Z_{A}(x))
\,, \label{eq:IR3}\end{eqnarray} 
where we dropped the Lorentz and group structure of the propagators. 
In \eq{eq:IR3}, $\Gamma_0^{(2)}= z\,x^\kappa=\hat z p^{2\kappa}$ are
the leading infrared terms for $k=0$ with $k$-independent prefactor
$\hat z$. The functions $\delta Z$ entail the transition between the
physical infrared regime, $x\gg 1$, and the cutoff regime, $x\ll1$. 
In \eq{eq:IR3} we have also used that non-renormalisation 
of the ghost-gluon vertex at
vanishing $k$ entails $\kappa_A=-2\kappa_C$ and
$\alpha_s=g^2/(4\pi\, z_Az_C^2)$ \cite{Lerche:2002ep}. 
Inserting the propagators \eq{eq:IR3} in \eq{eq:flow} leads to
two integral equations for $\delta Z_{A}$ and $\delta Z_C$ of the form
\begin{eqnarray}\label{eq:fixed}
\delta Z_{A/C}(x)=F_{A/C}[\delta Z_{A/C},\kappa_{C},\alpha_s]\,.
\end{eqnarray}
Explicit expressions for the integrals $F$ are given in
\cite{Pawlowski:2003hq}. The equations \eq{eq:fixed} are solved
iteratively for $\delta Z_{A}$, $\delta Z_C$, $\kappa_C$ and $
\alpha_s$, leading to 
\begin{eqnarray}\label{eq:result}
\kappa_C=0.59535\cdots\,,\qquad \qquad \alpha_s=2.9717\cdots\,.
\end{eqnarray}
The values in \eq{eq:result} are achieved by also invoking an
optimisation procedure developed in \cite{Pawlowski:2003hq} for
eliminating the regulator-dependence. 

The results \eq{eq:result} agree with the analytic results obtained
within DSEs \cite{Lerche:2002ep} and
stochastic quantisation \cite{Zwanziger:2001kw}. For the present
truncation within an optimised cut-off scheme one can indeed formally
link the integrated flow \eq{eq:flow} to a set of DSEs with explicit
renormalisation, see \cite{Pawlowski:2003hq}. We expect a mild $R$-dependence 
for the results as we have resorted to a truncation. Indeed, 
for general cut-off 
functions $R$, the results for $\kappa_C$ mildly vary in the interval 
\begin{eqnarray}\label{eq:interval}
\kappa_C\in [0.539\,,\,0.595]\,. 
\end{eqnarray} 
Both bounds have physical interpretations. The upper bound, as we
have already discussed, relates to the physical infinite volume result
whereas the lower bound can be linked to a finite volume computation.
The related regulator is a sharp cut-off that strictly allows no
propagation of modes with momenta $p^2$ smaller than $k^2$.  This is
as close to a finite volume (e.g.\ in lattice studies) as one can get 
with local momentum cut-off functions.
Interestingly, the lower bound $\kappa_C=0.539$ compares well to very
recent lattice results \cite{Oliveira:2004gy}.
A further interesting consequence of our analysis is the evaluation of
renormalisation procedures for DSEs stemming from the integrated flow
\eq{eq:flow}: for DSEs in the present truncation and subject to a
general consistent renormalisation it is impossible to achieve a
masslike behaviour for the gluon propagator, $\kappa_C=1/2$. More
generally $\kappa_C\in\!\!\!\!\!\!\!\slash\ [0,1/2]$, see 
\cite{Pawlowski:2003hq}. 
With multiplicative renormalisation this was already 
shown in \cite{Lerche:2002ep}. 

The above analysis allows for many interesting extensions. The most
important open question in the present truncation concerns the
detailed analysis of the transition between ultraviolet and infrared
regime. Unfortunately this question has not been resolved completely in
subsequent flow studies \cite{Kato:2004ry}. Moreover dynamical quarks
are investigated which opens the door towards a description of
dynamical chiral symmetry breaking.

\begin{theacknowledgments}
  JMP thanks the organisers of Quark Confinement and the Hadron
  Spectrum for making this interesting conference possible. This work
  has been supported by a grant from the Ministry of Science, Research
  and the Arts of Baden-W\"urttemberg (Az: 24-7532.23-19-18/1 and
  24-7532.23-19-18/2), EPSRC, and the DFG under contracts SM70/1-1 and 
GI328/1-2.

\end{theacknowledgments}

\end{document}